\documentclass[12pt,a4paper,dvips]{article}

\usepackage{a4p}
\usepackage{cite,mcite}
\usepackage{graphicx}
\usepackage{amssymb}
\usepackage{thisPhysics }
\usepackage{l3_titlePH,ifthen}

\journalname{Phys. Lett. B}
\preprint{2004-016}
\date{April 22, 2004}

\newlength{\capindent}
\newlength{\capwidth}
\newlength{\figwidth}
\newcommand{\icaption}[2][!*!,!]{\hspace*{\capindent}%
  \begin{minipage}{\capwidth}
    \ifthenelse{\equal{#1}{!*!,!}}%
      {\caption{#2}}%
      {\caption[#1]{#2}}
  \end{minipage}}

\def\tpi{\ensuremath{\mathrm{\tilde{\pi}}}}%
\def\fb{\mbox{fb$^{-1}$}}%
\begin{document}
\begin{titlepage}
%
\title{Search for Branons at LEP}
\author{The L3 Collaboration}
\begin{abstract}
We search, in the context of extra-dimension scenarios, for the
possible existence of brane fluctuations, called {\it branons}. Events
with a single photon or a single $\Zo$-boson and missing energy and momentum
collected with the L3 detector in $\epem$ collisions at centre-of-mass
energies $\rts=189-209 \GeV$ are analysed. No excess over the Standard
Model expectations is found and a lower limit at 95\% confidence level
of $103\GeV$ is derived for the mass of branons, for a scenario with
small brane tensions.  Alternatively, under the assumption of a light
branon, brane tensions below 180 GeV are excluded.
\end{abstract}

\submitted

\end{titlepage}


\section{Introduction}

The possible existence of additional space dimensions was suggested by
Kaluza and Klein \cite{KK} more than eighty years ago.  In the
original theory, the fundamental scale of gravitation, $M_F$,
coincides with the Planck scale, $M_P\approx 10^{19} \GeV$.  Since
then, several theories have used this idea as an alternative way to
solve some fundamental problems of physics, particularly those related with
gravitation and the unification of all forces. One of the most
attractive models is proposed in Reference \citen{ADD}. This model
assumes the restriction of the dynamics
of the Standard Model to a three-dimensional spatial brane, leaving
the gravitation and perhaps some other exotic particles the freedom
to propagate in the extra dimensions. If these extra dimensions have a
large size, $M_F$ is of the order of the electroweak scale and the
existence of extra dimensions could manifest at present and future
colliders with detection of gravitons, as described by an effective
theory with couplings of order $M_F$.

A different scenario is considered in this Letter. In this approach,
the presence of a three-dimensional brane as an additional physical
body in the theory, with its own dynamics, leads to the appearance of
additional degrees of freedom. These manifest as new scalar
particles, $\tpi$, called {\it branons}. Branons are
associated to brane fluctuations along the extra-dimensions
\cite{dobado_maroto} and are also natural dark-matter candidates
\cite{BCos}.  Their dynamics is determined by an effective theory with
couplings of the same order as the brane tension, $f$.

Searches for gravitons and branons are in a sense complementary
\cite{GB}.  If the brane tension is above the gravity scale, $f\gg
M_F$, the first evidence for extra dimensions would be the discovery
of gravitons, giving information about the fundamental scale of
gravitation and the characteristics of the extra dimensions.  If the
brane tension is below the gravity scale, $f\ll M_F$, then the first
signal of extra dimensions would be the discovery of branons, allowing
a measurement of the brane tension scale, the number of branons and
their masses \cite{ACDM}.

\begin{figure}[htbp]
  \begin{center}
    \includegraphics[width=\textwidth]{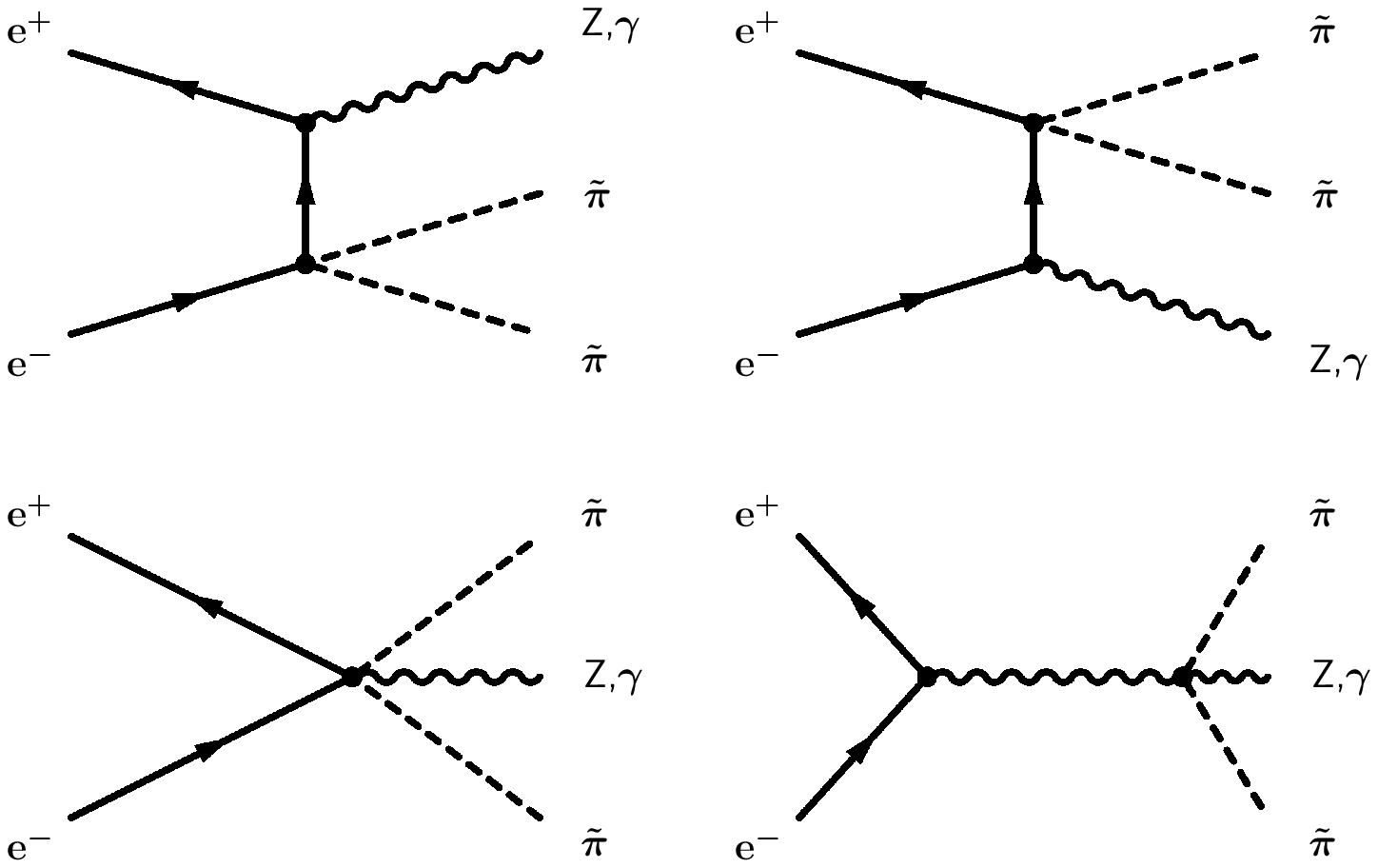}
  \end{center}
  \icaption{Feynman diagrams contributing to the branon production process: 
            $\ee\ra\tpi\tpi V_0$, where $V_0$ denotes a photon or a Z boson.\label{fig:branon_diagrams}}
\end{figure}

Many experimental results were reported on direct searches for
gravitons at LEP\cite{l3_sgpaper,l3_szpaper,lep_direct} and at the
TEVATRON~\cite{tevatron_direct}.  This Letter describes a search for
branons in data collected at LEP.  Branons couple to Standard Model
particles by pairs, suggesting the study of two production mechanisms
in $\rm e^+e^-$ collisions: $\ee\ra\tpi\tpi\gamma$ and
$\ee\ra\tpi\tpi\Zo$. They proceed via the diagrams shown in Figure
\ref{fig:branon_diagrams}. The experimental signature for branon
production at LEP is the presence of either a photon or a $\Zo$ boson
together with missing energy and momentum. This is due to the two branons which do
not interact in the detector and are hence invisible.  In the
following, only decays of the $\Zo$ boson into hadrons are
considered. For a given centre-of-mass energy, only the lighter
branons give a significant contribution to the cross sections of the
$\ee\ra\tpi\tpi\gamma$ and $\ee\ra\tpi\tpi\Zo$ processes. For
simplicity, we will assume a scenario with only one light branon
species of mass $M$.

%
\section{Data and Monte Carlo Samples}
%

Data collected by the L3 detector \cite{l3_det} at LEP in the
years from 1998 through 2000 are considered. They correspond to an
integrated luminosity of about 0.6~\fb\ at centre-of-mass energies,
$\sqrt{s}$, from 188.6 to 209.2 $\GeV$.

The following Monte Carlo generators are used to simulate Standard
Model processes: {\tt KK2f}~\cite{KKMC} for $\epem\ra
\nnbar\gam(\gam)$ and $\rm e^+ e^- \rightarrow q \overline q
(\gamma)$, {\tt GGG}~\cite{GGG} for \epem\ \ra\ \gam\gam(\gam), {\tt
BHWIDE}~\cite{BHWIDE} and {\tt TEEGG}~\cite{teegg} for large- and
small-angle Bhabha scattering, respectively, {\tt PHOJET}~\cite{phojet} and
{\tt DIAG36}~\cite{DIAG} for hadron and lepton production in
two-photon interactions, {\tt KORALW}~\cite{koralw} for W-boson
pair-production and {\tt EXCALIBUR}~\cite{excalibur_new} for Z-boson
pair-production and other four-fermion final states.  The predictions
of {\tt KK2f} for the $\epem\ \ra\ \nnbar\gam(\gam)$ process are
checked with the {\tt NUNUGPV}~\cite{NUNUGPV} generator.

The efficiencies for branon production through the processes
$\rm\ee\ra\tpi\tpi\gamma$ and $\rm\ee\ra\tpi\tpi\Zo\ra\tpi\tpi q
\overline q$ are determined by reweighting Monte Carlo events of the
processes $\epem \ra \nnbar\gam(\gam)$ and $\ee\ra\nnbar\Zo\ra\nnbar
\rm q\overline q$, respectively, with the differential cross sections of
Reference \citen{ACDM}. Events from the first process are generated
with {\tt KK2f} and events from the second process with {\tt
EXCALIBUR}, through W-boson fusion.

The L3 detector response is simulated using the {\tt GEANT}
program~\cite{geant}, which describes effects of energy loss, multiple
scattering and showering in the detector.  Time-dependent detector
inefficiencies, as monitored during the data-taking period, are
included in the simulation.

%

\section{Search in the \boldmath{$\rm\ee\ra\tpi\tpi\Zo\ra\tpi\tpi q \overline q$}  channel}

The single-Z signature for branon production at LEP is similar to the
signature of the associated production of a Z boson and a graviton which
was previously studied in data collected by L3 at $\rts=188.6
\GeV$\cite{l3_szpaper}. The events selected for that search are
re-analysed in this Letter to search for branons, and
the same analysis procedure is used to select candidate events at
$\rts=191.6-209.2 \GeV$. The integrated luminosities considered for
each value of $\rts$ are listed in Table~\ref{tab:sz_numbers}.

Unbalanced hadronic events with a visible mass compatible with that of
the Z boson are selected. The large background from $\rm e^+ e^-
\rightarrow q \overline q \gamma$ events with a low-angle high-energy
photon is reduced by requiring the missing momentum vector to point in
the detector. Cuts on event-shape and jet-shape variables are applied
to suppress other backgrounds: Z-boson pair-production with one of
the Z bosons decaying into neutrinos and the other into hadrons, 
W-boson pair-production with one of the W bosons decaying into hadrons
and the other into a low-angle undetected charged lepton and a
neutrino, and single-W production through the $\epem\ra\W \e \nu$
process, followed by a hadronic decay of the W boson.

Table \ref{tab:sz_numbers} summarises the yield of the selection at
the different centre-of-mass energies. In total, 455 events are
observed while 470 events are expected from several Standard Model
processes. The dominant background is W-boson pair-production (47\%).
Other sources of background are single-W production
(25\%), Z-boson pair-production (13\%) and the $\rm e^+ e^-
\rightarrow q \overline q \gamma$ process (12\%).

\begin{figure}[htbp]
  \begin{center}
    \includegraphics[height=0.4\textheight]{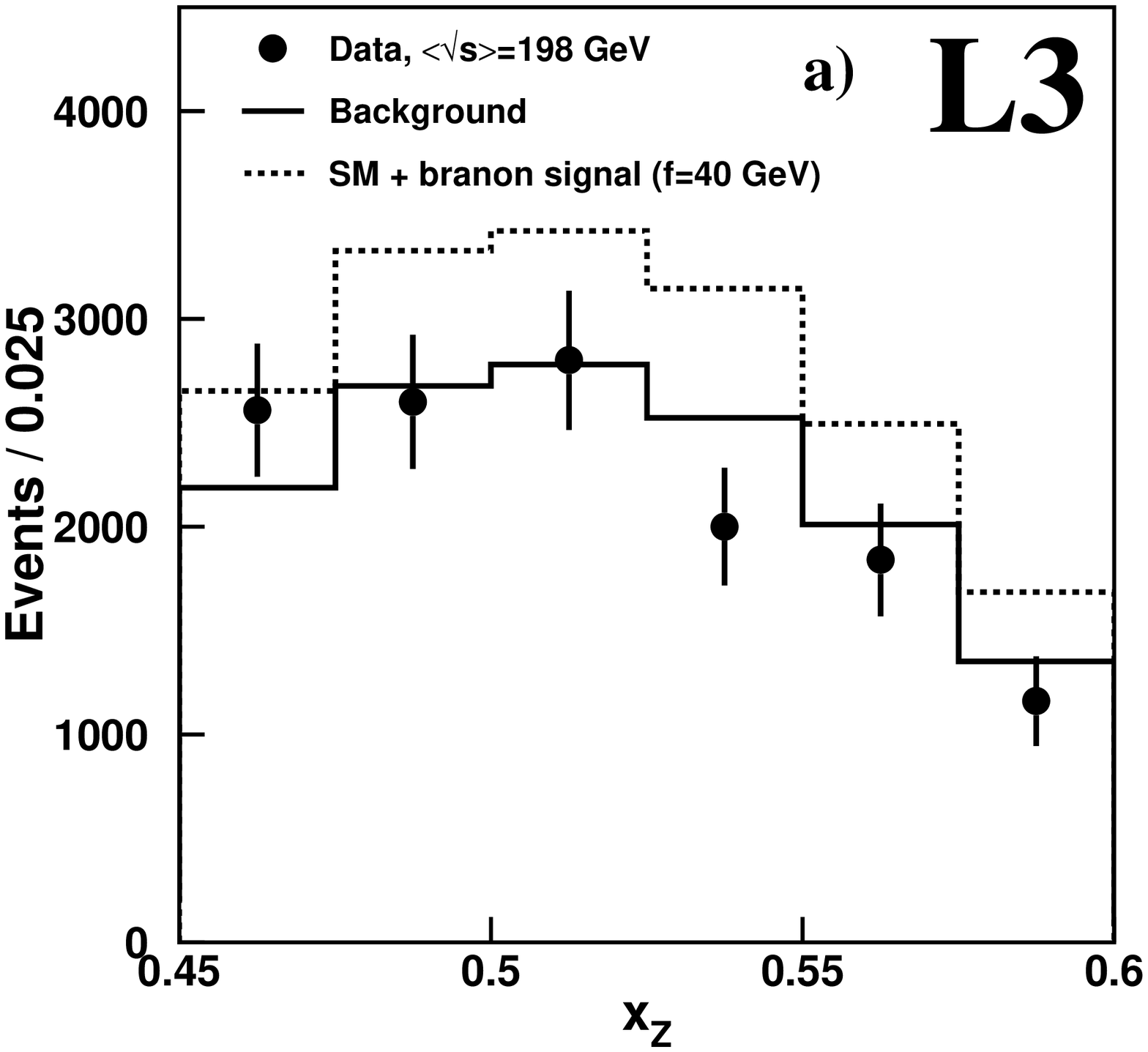}\\
    \includegraphics[height=0.4\textheight]{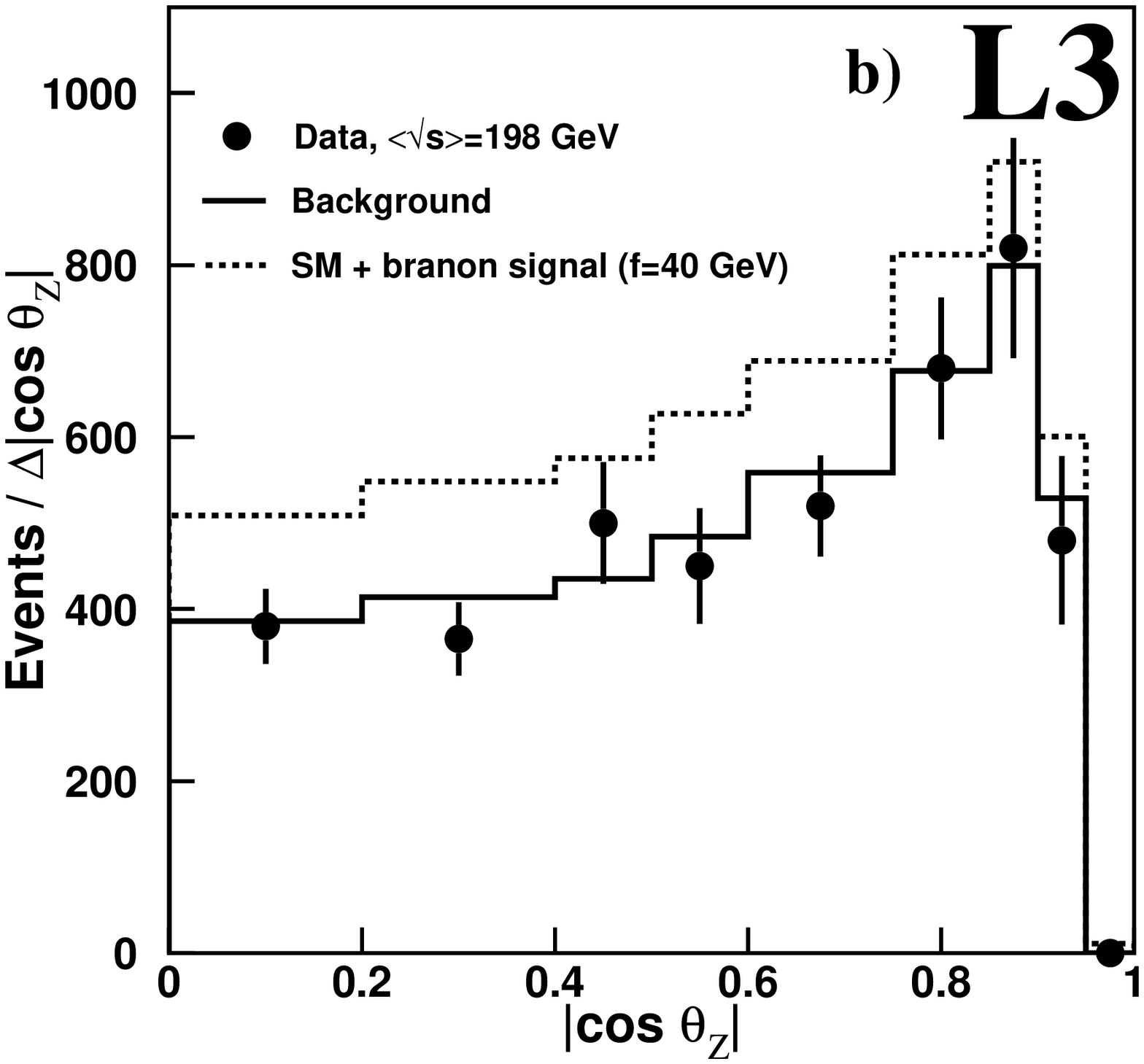}
  \end{center}
  \icaption{Distributions for events selected in the
	$\ee\ra\tpi\tpi\Zo\ra\tpi\tpi\rm q\overline q$ search: a)
	the reduced energy of the Z boson, $x_{\rm Z}=E_{\rm
	Z}/\sqrt{s}$, and b) the absolute value of the cosine of its
	polar angle, $\theta_{\rm Z}$.  The points represent the data, the
	solid histogram is the expectation from Standard Model
	processes and the dashed histogram is an example of the
	expectations in the presence of an additional signal due to
	branon production with $M=0$ and $f=40 \GeV$.
      \label{fig:plots_sz}
   }
\end{figure}

Expectations for a branon signal with $M=0$ and $f=40 \GeV$ are also
listed in Table \ref{tab:sz_numbers}.  The efficiency to detect such a
signal is 55\%. Two variables are most sensitive to discriminate a
possible signal from the Standard Model background: the reduced
energy of the Z boson, $x_{\rm Z}=E_{\rm Z}/\sqrt{s}$, and the cosine
of its polar angle, $\cos\theta_{\rm Z}$. The distribution of these
variables for data and Standard Model backgrounds are shown in Figures
\ref{fig:plots_sz}a and \ref{fig:plots_sz}b.  These Figures also show
the predictions in the presence of a branon signal. No excess with
respect to the Standard Model expectations is observed.


\section{Search in the \boldmath{$\ee\ra\tpi\tpi\gamma$} channel}

Events with a single photon and large missing energy and
momentum, selected by L3 at $\rts=188.6-209.2 \GeV$ \cite{l3_sgpaper},
are re-analysed for the presence of a signal due to the
$\ee\ra\tpi\tpi\gamma$ process in addition to the Standard Model
contributions from the $\ee\ra\nnbar\gamma(\gamma)$ and $\ee\ra\ee\gamma(\gamma)$ processes. A
breakdown of the integrated luminosities as a function of $\rts$ is
given in Table~\ref{tab:sg_numbers}.  Two different energy regimes
are considered, depending on the value of the transverse momentum of the
photon, $p_t$, relative to the beam energy, $E_{beam}$, and its polar angle, $\theta_\gamma$.  High-$p_t$
events, $0.04 E_{beam} < p_t < 0.60 E_{beam}$, are selected in both
the barrel, $|\cos\theta_\gamma|< 0.73$, and endcap, $0.8 <
|\cos\theta_\gamma| < 0.97$, regions of the electromagnetic
calorimeter. The selection of low-$p_t$ events, $0.016 E_{beam} < p_t
< 0.04 E_{beam}$, relies on a single-photon energy trigger with a
threshold around $900 \MeV$ which is active only in the barrel region
~\cite{bizzarri}.

\begin{figure}[htbp]
  \begin{center}
    \includegraphics[height=0.38\textheight]{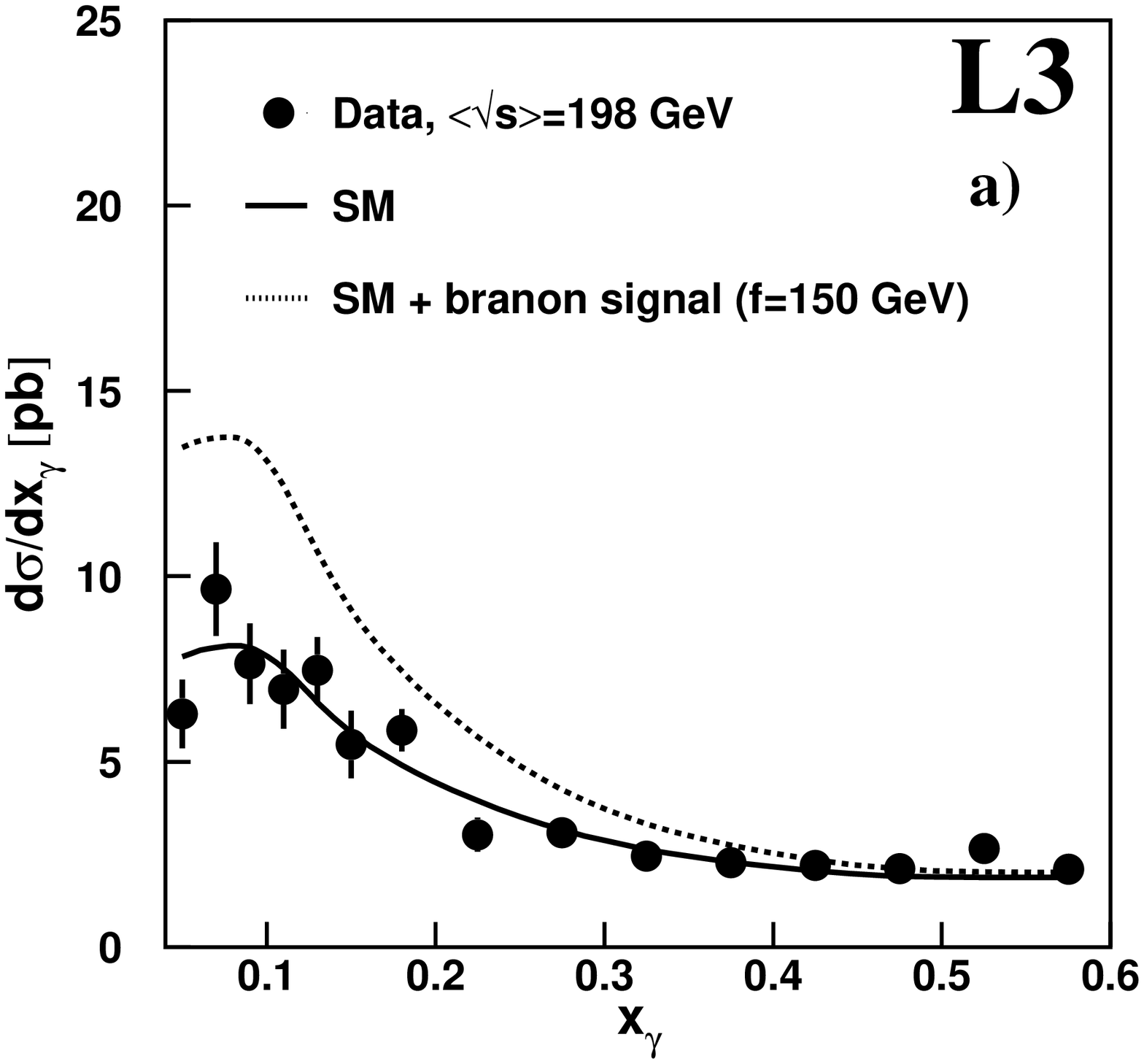} \\
    \includegraphics[height=0.38\textheight]{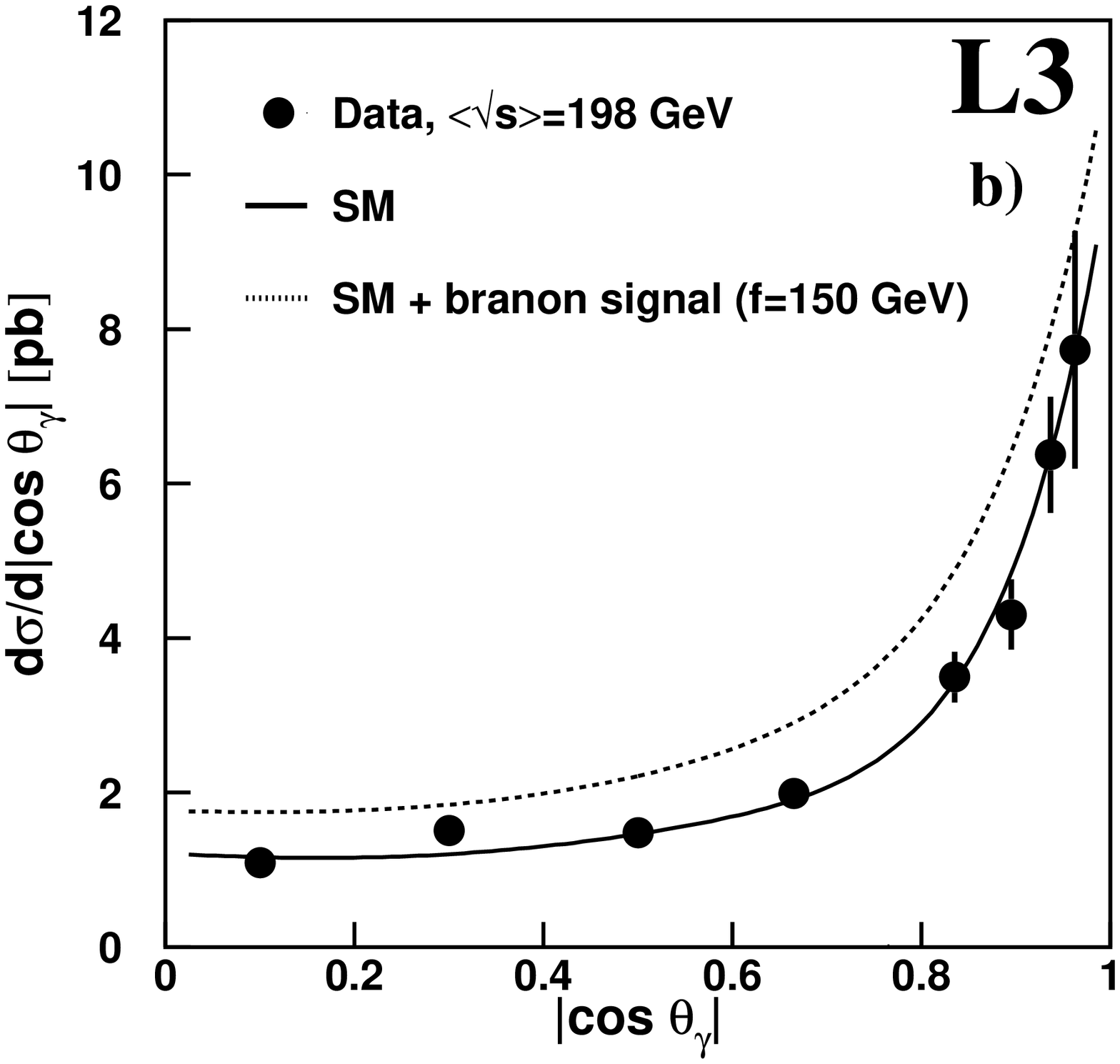}
  \end{center}
  \icaption{Measured differential cross sections for the
       $\ee\ra\nnbar\gamma(\gamma)$ process as a function of a)
       $x_\gamma=E_\gamma/E_{beam}$, the fraction of the beam energy
       carried by the photon and b) the absolute value of the cosine 
       of its polar angle, $\theta_\gamma$. Data
       selected by the high-$p_t$ selection at $0.04 E_{beam}<p_t<0.6
       E_{beam}$ are shown. They are integrated over the fiducial
       region $|\cos\theta_\gamma|<0.97$. The points represent the
       data, the solid curves are the Standard Model predictions and
       the dashed curves show the expectations in the presence of an
       additional signal due to branon production with $M=0$
       and $f=150 \GeV$.
      \label{fig:ene_g}
   }
\end{figure}

Table~\ref{tab:sg_numbers} lists the number of observed data events
together with the Standard Model expectations for different values of
$\rts$. The high-$p_t$ analysis selects $\ee\ra\nnbar\gamma(\gamma)$
events with purity above 99\% and efficiency above 80\%. In total,
838 events are observed in data while 811 are expected from Standard
Model processes.  Figures \ref{fig:ene_g}a and \ref{fig:ene_g}b show
the measured differential cross sections for the
$\ee\ra\nnbar\gamma(\gamma)$ processes as a function of
$x_\gamma=E_\gamma/E_{beam}$, the fraction of the beam energy carried
by the photon and of $|\cos{\theta_\gamma}|$.  Data obtained by the
high-$p_t$ selection are corrected for detector acceptance and
integrated over the polar-angle fiducial region
$|\cos\theta_\gamma|<0.97$. The measured differential cross sections 
are in good agreement with the Standard Model expectations.

\begin{figure}[htbp]
  \begin{center}
    \includegraphics[height=0.4\textheight]{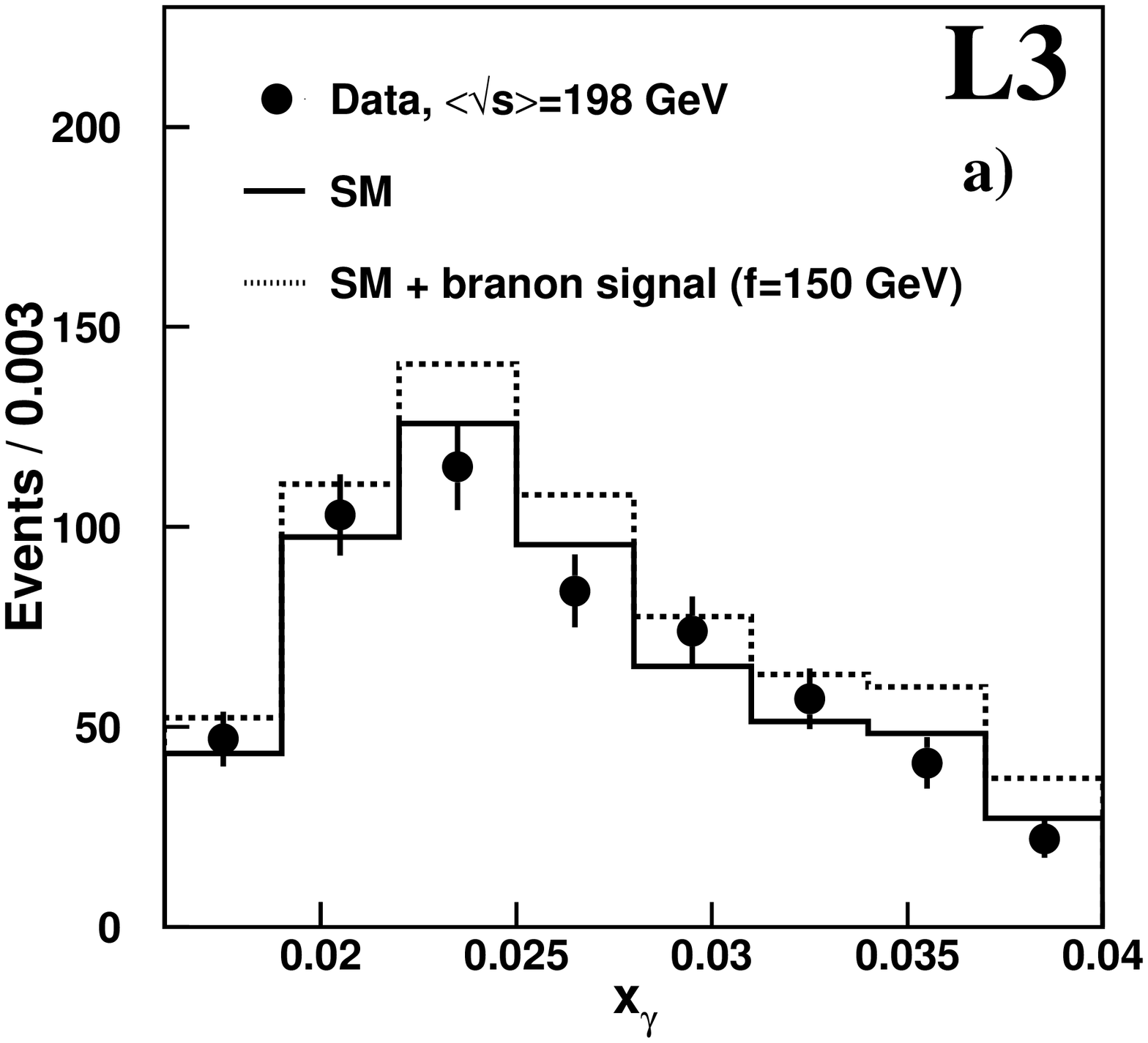} \\
    \includegraphics[height=0.4\textheight]{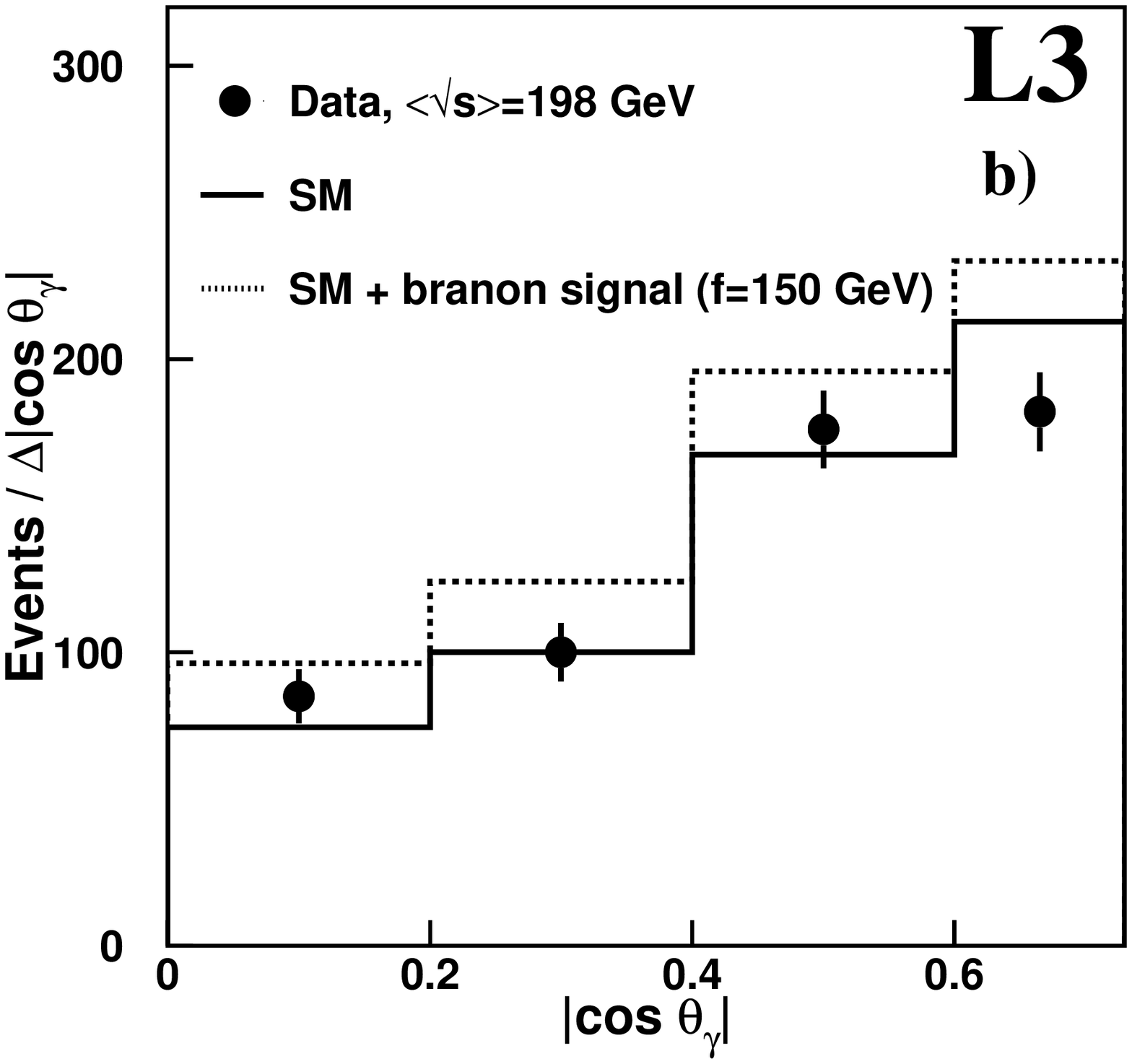}
  \end{center}
  \icaption{Distributions for events selected by the low-$p_t$
	selection at $0.016 E_{beam} < p_t < 0.04 E_{beam}$ in the
        $\ee\ra\tpi\tpi\gamma$ search: a) the fraction of the beam
        energy carried by the photon, $x_\gamma=E_\gamma/E_{beam}$ and
        b) the absolute value of the cosine of its polar angle,
        $\theta_\gamma$.  The points represent the data, the solid
        histogram is the expectation from Standard Model processes and
        the dashed histogram is an example of the expectations in the
        presence of an additional signal due to branon production with
        $M=0$ and $f=150 \GeV$.
      \label{fig:cos_g}
      }
\end{figure}

The criteria of the low-$p_t$ selections are much more stringent in
order to be sensitive to very low photon energies while minimizing the
huge $\ee\ra\ee\gamma(\gamma)$ component. In total, 543 events are
observed in data and 554 are expected from Standard Model
processes. The main contribution is from $\ee\ra\ee\gamma(\gamma)$
events and the $\ee\ra\nnbar\gamma(\gamma)$ purity is around $24\%$.
The event selection is described in detail in Reference
\citen{l3_sgpaper}.  Figures \ref{fig:cos_g}a and \ref{fig:cos_g}b
compare the distributions of $x_\gamma$ and $|\cos{\theta_\gamma}|$
observed in data with the expectations of the Standard Model
processes. A good agreement is observed.

The presence of a branon leads to an increase in the differential
cross sections which is a function of the branon mass $M$ and the
brane tension $f$ \cite{ACDM}:
\begin{eqnarray}
 \frac{{\rm d^2}\sigma(\ee\ra\tpi\tpi\gamma)}{{\rm d}x_\gamma {\rm d}\cos\theta_\gamma}&=&
           \alpha \frac{s(s(1-x_\gamma)-4M^2)^2}{61440 f^8\pi^2}
   \sqrt{1-\frac{4M^2}{s(1-x)}}
\nonumber\\
&&[x_\gamma(3-3x_\gamma+2x_\gamma^2)-x_\gamma^3\sin^2\theta_\gamma+
\frac{2(1-x_\gamma)(1+(1-x_\gamma)^2)}{x_\gamma\sin^2\theta_\gamma}],
\end{eqnarray}
where $\alpha$ is the electromagnetic coupling constant.

Figures \ref{fig:ene_g} and \ref{fig:cos_g} show the typical distortion in
the differential cross sections expected in the presence of a branon
signal.

%

\section{Results}

Evidence for branon production was found neither in the
$\rm\ee\ra\tpi\tpi\Zo\ra\tpi\tpi q \overline q$ nor in the
$\ee\ra\tpi\tpi\gamma$ channels and the data are interpreted in terms
of bounds on the possible production of branons. For each
centre-of-mass energy, the data and the expectations are compared in
bins of the two-dimensional distributions of $x_{\rm Z}$ {\it vs.}
$\cos{\theta_{\rm Z}}$ for the $\rm\ee\ra\tpi\tpi\Zo\ra\tpi\tpi q
\overline q$ channel and of $x_\gamma$ {\it vs.} $\cos{\theta_\gamma}$
for the $\ee\ra\tpi\tpi\gamma$ channel.  Assuming a Poisson
probability distribution for each bin, 95\% confidence level exclusion
limits are derived according to the method described in Reference
\citen{Agostini}. Systematic uncertainties are taken into account in
the calculation of the limit. For the $\rm\ee\ra\tpi\tpi\Zo\ra\tpi\tpi
q \overline q$ channel, they are similar to those encountered in the
study of Z-boson pair-production when one of the bosons decays into
hadrons and the other into neutrinos\cite{zz_paper} and are
dominated by uncertainties on the background normalisation, on the
detector energy scale and modelling and from limited Monte Carlo
statistics. The main systematic uncertainties for the
$\ee\ra\tpi\tpi\gamma$ channel\cite{l3_sgpaper} are 
the modelling of Standard Model process, the determination of
the trigger efficiency and the treatment of photons which convert in
electron-positron pairs in the detector material in front of the
electromagnetic calorimeter.

\begin{figure}[htbp]
  \begin{center}
    \includegraphics[width=\textwidth]{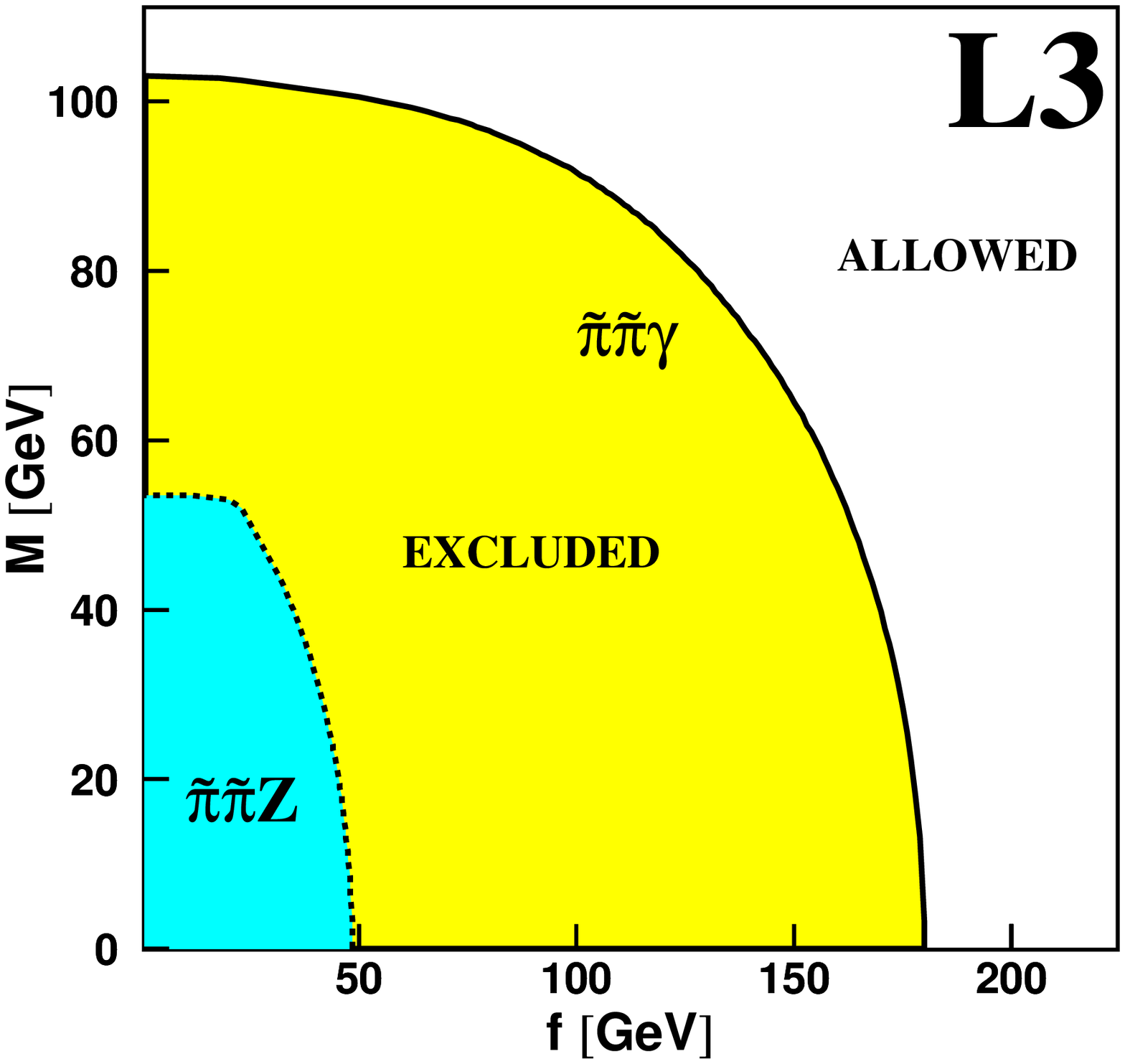}
  \end{center}
  \icaption{Two-dimensional regions in the $(f,M)$ plane excluded by
      the searches for branons produced in the $\epem\ra\tpi\tpi\Zo$
      and $\epem\ra\tpi\tpi\gamma$ processes. For very elastic branes, $f\ra
      0$, branon masses below $103 \GeV$ are excluded at 95\% confidence
      level. For massless branons, brane tensions below $180 \GeV$ are
      excluded.
      \label{fig:lim2}
  }
\end{figure}

The bounds from the $\ee\ra\tpi\tpi\Zo$ analysis are shown in Figure
\ref{fig:lim2}. For massless branons, the brane tension $f$ must be
greater than $47\GeV$. There is no sensitivity for branon masses near
and beyond the kinematic limit ($M\gtrsim (\rts-\MZ)/2$) and no bounds
on $f$ can be derived for $M>54 \GeV$.  The sensitivity in the
$\ee\ra\tpi\tpi\gamma$ channel is larger than that of the
$\rm\ee\ra\tpi\tpi\Zo\ra\tpi\tpi q \overline q$ channel. This is due
to two factors: the different coupling of the Z boson and the photon
to electrons and a larger phase space available in the presence of a
photon in the final state, as opposed to a massive Z boson.  The
limits obtained from the $\ee\ra\tpi\tpi\gamma$ analysis are also
shown in Figure \ref{fig:lim2}.  For $M=0$, the brane tension $f$ must
be greater than 180 GeV. For very elastic branes ($f\ra 0$) a lower
branon mass bound of $M>103 \GeV$ is obtained.

These bounds are the most stringent to date on the possible existence
of branons.  The bounds for $M>0$ GeV complement and improve those
deduced from astrophysical observations~\cite{BCos}.

\bibliographystyle{l3style}
\bibliography{branon}

%
%

\newpage
\typeout{   }     
\typeout{Using author list for paper 287 -  }
\typeout{$Modified: Jul 15 2001 by smele $}
\typeout{!!!!  This should only be used with document option a4p!!!!}
\typeout{   }
%
%
%
%
%
%

\newcount\tutecount  \tutecount=0
\def\tutenum#1{\global\advance\tutecount by 1 \xdef#1{\the\tutecount}}
\def\tute#1{$^{#1}$}
\tutenum\aachen            
\tutenum\nikhef            
\tutenum\mich              
\tutenum\lapp              
\tutenum\basel             
\tutenum\lsu               
\tutenum\beijing           
\tutenum\bologna           
\tutenum\tata              
\tutenum\ne                
\tutenum\bucharest         
\tutenum\budapest          
\tutenum\mit               
\tutenum\panjab            
\tutenum\debrecen          
\tutenum\dublin            
\tutenum\florence          
\tutenum\cern              
\tutenum\wl                
\tutenum\geneva            
\tutenum\hamburg           
\tutenum\hefei             
\tutenum\lausanne          
\tutenum\lyon              
\tutenum\madrid            
\tutenum\complutense	   
\tutenum\florida           
\tutenum\milan             
\tutenum\moscow            
\tutenum\naples            
\tutenum\cyprus            
\tutenum\nymegen           
\tutenum\caltech           
\tutenum\perugia           
\tutenum\peters            
\tutenum\cmu               
\tutenum\potenza           
\tutenum\prince            
\tutenum\riverside         
\tutenum\rome              
\tutenum\salerno           
\tutenum\ucsd              
\tutenum\sofia             
\tutenum\korea             
\tutenum\taiwan            
\tutenum\tsinghua          
\tutenum\purdue            
\tutenum\psinst            
\tutenum\zeuthen           
\tutenum\eth               

{
\parskip=0pt
\noindent
{\bf The L3 Collaboration:}
\ifx\selectfont\undefined
 \baselineskip=10.8pt
 \baselineskip\baselinestretch\baselineskip
 \normalbaselineskip\baselineskip
 \ixpt
\else
 \fontsize{9}{10.8pt}\selectfont
\fi
\medskip
\tolerance=10000
\hbadness=5000
\raggedright
\hsize=162truemm\hoffset=0mm
\def\r{\rlap,}
\noindent

P.Achard\r\tute\geneva\ 
O.Adriani\r\tute{\florence}\ 
M.Aguilar-Benitez\r\tute\madrid\ 
J.Alcaraz\r\tute{\madrid}\ 
G.Alemanni\r\tute\lausanne\
J.Allaby\r\tute\cern\
A.Aloisio\r\tute\naples\ 
M.G.Alviggi\r\tute\naples\
H.Anderhub\r\tute\eth\ 
V.P.Andreev\r\tute{\lsu,\peters}\
% ...
\newpage
\begin{list}{A}{\itemsep=0pt plus 0pt minus 0pt\parsep=0pt plus 0pt minus 0pt
                \topsep=0pt plus 0pt minus 0pt}
\item[\aachen]
 III. Physikalisches Institut, RWTH, D-52056 Aachen, Germany$^{\S}$
\item[\nikhef] National Institute for High Energy Physics, NIKHEF, 
     and University of Amsterdam, NL-1009 DB Amsterdam, The Netherlands
\item[\mich] University of Michigan, Ann Arbor, MI 48109, USA
\item[\lapp] Laboratoire d'Annecy-le-Vieux de Physique des Particules, 
     LAPP,IN2P3-CNRS, BP 110, F-74941 Annecy-le-Vieux CEDEX, France
\item[\basel] Institute of Physics, University of Basel, CH-4056 Basel,
     Switzerland
\item[\lsu] Louisiana State University, Baton Rouge, LA 70803, USA
\item[\beijing] Institute of High Energy Physics, IHEP, 
  100039 Beijing, China$^{\triangle}$ 
\item[\bologna] University of Bologna and INFN-Sezione di Bologna, 
     I-40126 Bologna, Italy
\item[\tata] Tata Institute of Fundamental Research, Mumbai (Bombay) 400 005, India
\item[\ne] Northeastern University, Boston, MA 02115, USA
\item[\bucharest] Institute of Atomic Physics and University of Bucharest,
     R-76900 Bucharest, Romania
\item[\budapest] Central Research Institute for Physics of the 
     Hungarian Academy of Sciences, H-1525 Budapest 114, Hungary$^{\ddag}$
\item[\mit] Massachusetts Institute of Technology, Cambridge, MA 02139, USA
\item[\panjab] Panjab University, Chandigarh 160 014, India
\item[\debrecen] KLTE-ATOMKI, H-4010 Debrecen, Hungary$^\P$
\item[\dublin] Department of Experimental Physics,
  University College Dublin, Belfield, Dublin 4, Ireland
\item[\florence] INFN Sezione di Firenze and University of Florence, 
     I-50125 Florence, Italy
\item[\cern] European Laboratory for Particle Physics, CERN, 
     CH-1211 Geneva 23, Switzerland
\item[\wl] World Laboratory, FBLJA  Project, CH-1211 Geneva 23, Switzerland
\item[\geneva] University of Geneva, CH-1211 Geneva 4, Switzerland
\item[\hamburg] University of Hamburg, D-22761 Hamburg, Germany
\item[\hefei] Chinese University of Science and Technology, USTC,
      Hefei, Anhui 230 029, China$^{\triangle}$
\item[\lausanne] University of Lausanne, CH-1015 Lausanne, Switzerland
\item[\lyon] Institut de Physique Nucl\'eaire de Lyon, 
     IN2P3-CNRS,Universit\'e Claude Bernard, 
     F-69622 Villeurbanne, France
\item[\madrid] Centro de Investigaciones Energ{\'e}ticas, 
     Medioambientales y Tecnol\'ogicas, CIEMAT, E-28040 Madrid,
     Spain${\flat}$ 
\item[\complutense] Universidad Complutense de Madrid, E-28040 Madrid,
     Spain
\item[\florida] Florida Institute of Technology, Melbourne, FL 32901, USA
\item[\milan] INFN-Sezione di Milano, I-20133 Milan, Italy
\item[\moscow] Institute of Theoretical and Experimental Physics, ITEP, 
     Moscow, Russia
\item[\naples] INFN-Sezione di Napoli and University of Naples, 
     I-80125 Naples, Italy
\item[\cyprus] Department of Physics, University of Cyprus,
     Nicosia, Cyprus
\item[\nymegen] University of Nijmegen and NIKHEF, 
     NL-6525 ED Nijmegen, The Netherlands
\item[\caltech] California Institute of Technology, Pasadena, CA 91125, USA
\item[\perugia] INFN-Sezione di Perugia and Universit\`a Degli 
     Studi di Perugia, I-06100 Perugia, Italy   
\item[\peters] Nuclear Physics Institute, St. Petersburg, Russia
\item[\cmu] Carnegie Mellon University, Pittsburgh, PA 15213, USA
\item[\potenza] INFN-Sezione di Napoli and University of Potenza, 
     I-85100 Potenza, Italy
\item[\prince] Princeton University, Princeton, NJ 08544, USA
\item[\riverside] University of Californa, Riverside, CA 92521, USA
\item[\rome] INFN-Sezione di Roma and University of Rome, ``La Sapienza",
     I-00185 Rome, Italy
\item[\salerno] University and INFN, Salerno, I-84100 Salerno, Italy
\item[\ucsd] University of California, San Diego, CA 92093, USA
\item[\sofia] Bulgarian Academy of Sciences, Central Lab.~of 
     Mechatronics and Instrumentation, BU-1113 Sofia, Bulgaria
\item[\korea]  The Center for High Energy Physics, 
     Kyungpook National University, 702-701 Taegu, Republic of Korea
\item[\taiwan] National Central University, Chung-Li, Taiwan, China
\item[\tsinghua] Department of Physics, National Tsing Hua University,
      Taiwan, China
\item[\purdue] Purdue University, West Lafayette, IN 47907, USA
\item[\psinst] Paul Scherrer Institut, PSI, CH-5232 Villigen, Switzerland
\item[\zeuthen] DESY, D-15738 Zeuthen, Germany
\item[\eth] Eidgen\"ossische Technische Hochschule, ETH Z\"urich,
     CH-8093 Z\"urich, Switzerland
\item[\S]  Supported by the German Bundesministerium 
        f\"ur Bildung, Wissenschaft, Forschung und Technologie.
\item[\ddag] Supported by the Hungarian OTKA fund under contract
numbers T019181, F023259 and T037350.
\item[\P] Also supported by the Hungarian OTKA fund under contract
  number T026178.
\item[$\flat$] Supported also by the Comisi\'on Interministerial de Ciencia y 
        Tecnolog{\'\i}a.
\item[$\sharp$] Also supported by CONICET and Universidad Nacional de La Plata,
        CC 67, 1900 La Plata, Argentina.
\item[$\triangle$] Supported by the National Natural Science
  Foundation of China.
\end{list}
}
\vfill


\newpage

%
%

\begin{table}
\begin{center}
\begin{tabular}{|c|c|c|c|c|}                                       
      \hline
      $\sqrt{s}$ (\GeV{}) & ${\cal L}$ (pb$^{-1}$) & $N_D$ & $N_B$ & $N_S$ \\ 
      \hline
      188.6         &           176.8&            129 &           126.3 & \phantom{1}20.9\\
      191.6         & \phantom{1}29.8&  \phantom{1}22 & \phantom{1}23.1 & \phantom{11}4.1\\
      195.5         & \phantom{1}84.1&  \phantom{1}63 & \phantom{1}65.7 & \phantom{1}12.8\\
      199.5         & \phantom{1}83.3&  \phantom{1}59 & \phantom{1}63.7 & \phantom{1}18.0\\
      201.7         & \phantom{1}37.2&  \phantom{1}30 & \phantom{1}28.9 & \phantom{11}8.8\\
      202.5$-$205.5 & \phantom{1}79.0&  \phantom{1}51 & \phantom{1}59.9 & \phantom{1}20.5\\
      205.5$-$209.2 &           139.0&            101 &           102.2 & \phantom{1}37.9\\
      \hline
      Total         &           629.2&            455 &           469.8 &           123.0\\
      \hline
\end{tabular}
\end{center}
      \icaption{Luminosity, number of selected data events, $N_D$, and expected
	Standard Model background events, $N_B$, as a function of $\rts$ in the search for the
	$\ee\ra\tpi\tpi\Zo\ra\tpi\tpi\rm q\overline q$ process. The
	expected number of signal events, $N_S$, is also given for $M=0$
	and $f=40 \GeV$.
        \label{tab:sz_numbers}}
\end{table}

\begin{table}
\begin{center}
\begin{tabular}{|c|c|c|c|c|c|c|c|}
      \cline{3-8}
      \multicolumn{2}{c|}{} & \multicolumn{3}{|c|}{High $p_t$}& \multicolumn{3}{|c|}{Low $p_t$}\\
      \hline
      $\sqrt{s}$ (\GeV{}) & ${\cal L}$ (pb$^{-1}$) & $N_D$ & $N_B$ & $N_S$& $N_D$ & $N_B$ & $N_S$\\ 
      \hline
      188.6         &           176.0 &           249 &           254.7 & \phantom{1}71.0  &           156 &           152.9 &           16.6\\
      191.6         & \phantom{1}29.5 & \phantom{1}37 & \phantom{1}40.2 & \phantom{1}13.0  & \phantom{1}32 & \phantom{1}28.2 & \phantom{1}3.5\\
      195.5         & \phantom{1}83.9 &           123 &           110.1 & \phantom{1}42.9  & \phantom{1}73 & \phantom{1}80.1 &           11.5\\
      199.5         & \phantom{1}81.3 &           114 &           102.6 & \phantom{1}47.2  & \phantom{1}74 & \phantom{1}77.0 &           13.0\\
      201.7         & \phantom{1}34.8 & \phantom{1}53 & \phantom{1}43.9 & \phantom{1}21.8  & \phantom{1}35 & \phantom{1}32.7 & \phantom{1}6.0\\
      202.5$-$205.5 & \phantom{1}74.8 &           103 & \phantom{1}90.4 & \phantom{1}52.3  & \phantom{1}71 & \phantom{1}70.6 &           15.1\\
      205.5$-$207.2 &           130.2 &           151 &           158.9 & \phantom{1}96.5  & \phantom{1}93 &           105.9 &           27.7\\
      207.2$-$209.2 & \phantom{11}8.6 & \phantom{11}8 & \phantom{1}10.4 & \phantom{11}6.7  & \phantom{11}9 & \phantom{11}7.0 & \phantom{1}1.9\\ 
      \hline
      Total         &           619.1 &           838 &           811.2 &           351.4  &           543 &           554.4 &           95.3\\ 
      \hline
\end{tabular}
\end{center}
      \icaption{Luminosity, number of selected data events, $N_D$, and
 expected Standard Model background events, $N_B$, as a function of
 $\rts$ for the high-$p_t$ and the low-$p_t$ single-photon
 selections. Expectations, $N_S$, for branon production through the
 $\ee\ra\tpi\tpi\gamma$ process are also given for $M=0$ and $f=150 \GeV$.
 \label{tab:sg_numbers}}
\end{table}

\newpage

%
%
  
\end{document}